\documentclass[prl,aps,floats,twocolumn]{revtex4}

\usepackage[dvips]{graphicx}
\usepackage{amssymb}
\usepackage{amsmath}

\usepackage{color}

\begin{document}

\title{Light Loop Echoes and Blinking Black Holes}

\author{Latham Boyle$^{1,2}$ and Matthew Russo$^{1}$}

\affiliation{$^{1}$Canadian Institute for Theoretical Astrophysics
  (CITA), Toronto, Ontario, Canada \\
  $^{2}$Perimeter Institute for Theoretical Physics, Waterloo, Ontario, Canada}

\date{October 2011}
                            
\begin{abstract}
  Radiation emitted near a black hole reaches the observer by multiple
  paths; and when this radiation varies in time, the time-delays
  between the various paths generate a ``blinking'' effect in the
  observed light curve $L(t)$ or its autocorrelation function
  $\xi(T)=\langle L(t)L(t-T)\rangle$.  For the particularly important
  ``face-on'' configuration (in which the hole is viewed roughly along
  its spin axis, while the emission comes roughly from its equatorial
  plane -- {\it e.g.}\ from the inner edge of its accretion disk, or
  from the violent flash of a nearby/infalling star) we calculate the
  blinking in detail by computing the time delay $\Delta
  t_{j}(r_{\ast},a)$ and magnification $\mu_{j}(r_{\ast},a)$ of the
  $j$th path $(j=1,2,3,\ldots)$, relative to the primary path ($j=0$),
  as a function of the emission radius $r_{\ast}$ and black hole spin
  $0\leq a/M\leq1$.  The particular geometry and symmetry of the
  nearly-face-on configuration enhances and ``protects'' the blinking
  signal, making it more detectable and more independent of certain
  astrophysical and observational details.  The effect can be
  surprisingly strong: {\it e.g.}\ for radiation from the innermost
  stable circular orbit (``ISCO'') of a black hole of critical spin
  ($a_{{\rm crit}}/M\approx0.853$), the $j=1,2,3$ fluxes are,
  respectively, $27\%$, $2\%$ and $0.1\%$ of the $j=0$ flux.
\end{abstract}
\maketitle 

Light rays are bent as they pass through curved regions of spacetime.
To date, physicists have only detected rays with tiny bending angles
($\ll\!2\pi$, even in the famous ``strong-lensing'' systems, where
galaxies appear to be stretched into banana-shaped arcs on the sky).
On the other hand, rays that pass very near a black hole can
experience large bending angles, and even be bent into ``light loops''
that circle the hole once or more before proceeding to the observer
\cite{Darwin} (see Fig.~1).  Detection of such highly bent rays would
provide an unprecedented test of strong-field general relativity, and
a precious new window onto the physics and astrophysics near black
holes.

Previous authors have suggested various ways to look for these light
loops observationally \cite{CunninghamBardeen, HolzWheeler, BozzaEtAl,
  BroderickLoeb}.  In this paper, we investigate a different strategy.
We start from the idea that the emission from an intrinsically
time-varying source very near a black hole will reach the observer by
multiple paths; and the time-delay between the different paths will
induce a characteristic ``blinking'' signal in the observed light
curve $L(t)$ or its auto-correlation function $\xi(T)=\langle
L(t)L(t-T)\rangle$.  From here, we are led to focus on the ``face-on''
or ``right-angle'' configuration (in which the hole is viewed roughly
along its spin axis, while the emission comes roughly from its
equatorial plane -- {\it e.g.}\ from the inner edge of its accretion
disk, or from the violent flash of a nearby/infalling star).  As we
shall explain, a variety of mathematical, astrophysical and
observational considerations point to this configuration as being of
special importance when it comes to detecting blinking black holes:
just as the nearly-straight-line configuration of Fig.~1a is the ideal
geometry for ordinary gravitational lensing, the nearly-face-on
configuration of Fig.~1b may be regarded as the ideal geometry for
blinking black holes.  For this configuration, we compute the blinking
signal in detail (by computing the time delay $\Delta t_{j}$ and
magnification $\mu_{j}$ of each light loop relative to the primary
light path) as a function of: (i) the distance $r_{\ast}$ between the
hole and the source, and (ii) the spin $0\leq a/M\leq1$ of the hole.
The blinking signal can be surprisingly strong, and we hope it may be
detectable.
\begin{figure}
  \begin{center}
    \includegraphics[width=3.1in]{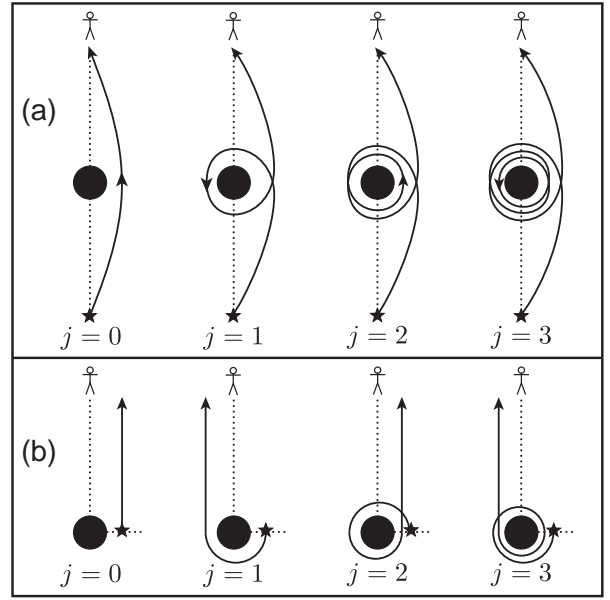}
  \end{center}
  \caption{Schematic light loops: (a) the standard straight-line
    configuration; (b) the right-angle/face-on configuration.}
  \label{blinkingFig}
\end{figure}

To see how we might try to detect light-loops, it is useful to start
by understanding why, at first glance, the task seems practically
impossible!  Consider the standard gravitational lensing
configuration, in which the lens is nearly aligned between the source
and the observer, and far away from both (Fig.~1a).  If the lens is a
non-spinning (Schwarzschild) black hole of mass $M$, the observer sees
an infinite series ($j=0,1,2,\ldots$) of concentric Einstein rings on
the sky \cite{Darwin}: the outer ($j=0$) ring is the ordinary one,
while the inner ($j\geq1$) rings are due to light loops with bending
angles $\alpha_{j}^{}\approx2\pi j$.  The $j\geq1$ rings are extremely
dim relative to the $j=0$ ring.  To see this, note that the bending
angle $\alpha$ depends on the impact parameter $b$ as:
$\alpha(b)\approx 4M/b$ (for small $\alpha$) and $\alpha(b)\approx
{\rm ln}[3.48M/(b-b_{{\rm cr}})]$ (for large $\alpha$) \cite{Darwin},
where $b_{{\rm cr}} =3\sqrt{3}M$.  Given this, standard lensing
analysis \cite{GravitationalLensing} implies that, in the limit of
perfect source/lens/observer alignment, the magnification $\mu_{j}$ of
the $j$th image ($j\geq1$) {\it relative to the $0$th image} (rather
than the unlensed image) is
\begin{equation}
  \label{standard_light_loops}
  \mu_{j}\approx 9[MD_{S}/D_{L}D_{LS}]^{3/2}{\rm e}^{-2\pi j}
\end{equation}
where $D_{S}$ is the observer-source distance, $D_{L}$ is the
observer-lens distance, and $D_{LS}$ is the lens-source distance.
This expression seems discouraging for two reasons: (i) the factor in
square brackets looks tiny because in ordinary lensing $D_{L}$,
$D_{S}$ and $D_{LS}$ are enormous relative to the Schwarzschild radius
$2M$ of the lens; and (ii) the factor ${\rm e}^{-2\pi j}$ says that to
see highly bent rays, we must pay an exponential price (as the bending
angle $\alpha$ increases, the magnification of the corresponding image
is suppressed by ${\rm e}^{-\alpha}$).  But before getting
discouraged, note that we can improve the situation dramatically via
the following two tricks.  First, if we bring the source near the
lens, so that $D_{LS}\sim M$, and hence $D_{L}\approx D_{S}$, then the
factor in square brackets will be ${\cal O}(1)$.  Second, if we switch
from the straight-line configuration of Fig.~1a to the right-angle
configuration of Fig.~1b, then instead of successive images being
suppressed by ${\rm e}^{-2\pi}\approx0.0019$, they are only suppressed
by ${\rm e}^{-\pi}\approx0.043$.

Nature may be kind enough to provide astronomical systems that take
advantage of these two tricks.  For example, a black hole is often
surrounded by an accretion disk whose inner edge \cite{PageThorne}
lies near the hole's innermost stable circular orbit or ``ISCO'' and
can be a strongly time-varying radiation source; furthermore, many
such holes (especially those that have grown significantly via
accretion) are thought to be rapidly spinning \cite{Bardeen:1970},
which brings the ISCO even closer to the hole (see Fig.~2a).  The
Bardeen-Petterson effect \cite{BardeenPetterson} tends to align the
inner accretion disk with the equatorial plane of a spinning hole, so
the right-angle configuration of Fig.~1b corresponds to viewing the
inner accretion disk nearly face-on, and nearly along the hole's spin
axis.  With this initial motivation, let us now do a full calculation
of the blinking signal generated by a spinning black hole in the
face-on geometry.  As we proceed, we will encounter a variety of other
reasons to be interested in this configuration.
\begin{figure}
  \begin{center}
    \includegraphics[width=3.1in]{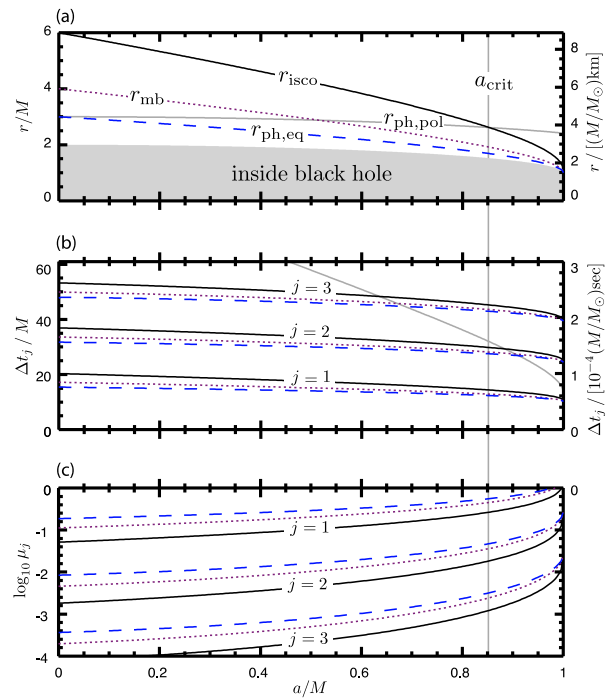}
  \end{center}
  \caption{Panel a: $r_{{\rm ph,eq}}$, $r_{{\rm mb}}$ and $r_{{\rm
        isco}}$ are, respectively, the innermost, innermost {\it
      bound}, and innermost {\it stable} circular orbits
    \cite{Darwin,BardeenPressTeukolsky}.  The polar ($L_{z}=0$) photon
    orbit has radius $r_{{\rm ph,pol}}$, which intersects $r_{{\rm
        isco}}$ at the critical spin $a_{{\rm crit}}\approx0.853$.
    Panels b,c: Time delay $\Delta t_{j}(r_{\ast},a)$ and
    magnification $\mu_{j}(r_{\ast},a)$ of the $j=1,2,3$ paths
    relative to the $j=0$ path, for $r_{\ast}=r_{\rm ph,eq}$ (dashed
    blue curve), $r_{\ast}=r_{\rm mb}$ (dotted red curve), and
    $r_{\ast}=r_{{\rm isco}}$ (solid black curve).  The solid grey
    curve in Panel b is the orbital period at $r_{{\rm isco}}$; note
    that it is considerably longer than the separation between
    blinks.}
 \label{rISCOcritFig}
\end{figure}

A spinning black hole of mass $M$ and spin $a$ is described by the
Kerr metric \cite{Kerr} (see \cite{Carter, BardeenPressTeukolsky,
  Chandrasekhar} for an introduction).  We use Boyer-Lindquist
coordinates and choose units with $G=c=M=1$ so that all quantities
become dimensionless and $0\leq a\leq1$.  Light rays are the null
geodesics of this metric.  Along any such geodesic $x^{\mu}(\lambda)$,
with tangent vector $p^{\mu}=dx^{\mu}/d\lambda$, there are 3 conserved
quantities: the energy $E=-p_{t}$, the axial angular momentum
$L_{z}=p_{\varphi}$, and the Carter constant $Q=p_{\theta}^{2}+{\rm
  cos}^{2}(\theta)[p_{\varphi}^{2}/{\rm sin}^{2}(\theta)
-a^{2}p_{t}^{2}]$.  We rescale the affine parameter $\lambda$ by a
constant so that $E=1$.  Let us start by imagining that a source in a
nearly circular equatorial orbit around the black hole emits a flash
that is isotropic in the rest frame of the source.  The null geodesics
connecting the flash at $(r=r_{\ast}, \theta=\pi/2)$ to the face-on
observer at $(r=\infty,\theta=0)$ form an infinite series labeled by a
non-negative integer ($j=0,1,2,\ldots$).  Along the $j$th geodesic the
polar angle $\theta$ varies by a total amount
$\int|d\theta|=(2j+1)\pi/2$, as shown in Fig.~1b; the azimuthal angle
also varies ($\int|d\varphi|\neq0$), but we do not need to compute
this variation in order to predict the blinking signal in the face-on
limit.  The $j$th geodesic is characterized by vanishing axial angular
momentum $L_{z}=0$, and a positive Carter constant
$Q_{j}(r_{\ast},a)>0$, which is determined by the requirement that $r$
and $\theta$ obey the relevant first integral of the geodesic equation
\cite{Carter, Chandrasekhar}
\begin{equation}
  \frac{|d\theta|}{\!\!\sqrt{Q\!+\!a^{2}{\rm cos}^{2}\theta}}\!=\!
  \frac{|dr|}{\!\!\sqrt{r^{4}\!+\!(a^{2}\!-\!Q)r^{2}\!+\!2(a^{2}\!+\!Q)r\!-\!a^{2}Q}}
\end{equation}
as well as the boundary conditions described above.  In practice, we
must solve for $Q_{j}(r_{\ast},a)$ numerically.  In doing so, note
that when $j\geq1$ and $r_{\ast}$ is sufficiently large, the geodesic
initially heads inward ($dr/dt<0$), reaches a radial turning point at
$R(r)=0$, and then heads outward ($dr/dt>0$) to the observer.

Given $Q_{j}(r_{\ast},a)$, we use Eqs.~(180,185,186) in Sec.~62 of
\cite{Chandrasekhar} to find the observed time delay $\Delta
t_{j}(r_{\ast},a)=t_{j}-t_{0}$ between the $j$th and $0th$ flashes.
We can also use $Q_{j}(r_{\ast},a)$ to compute $\mu_{j}$, the ratio
between the observed energy flux in the $j$th flash and the $0$th
flash, as follows.  The observed energy flux in the $j$th flash is the
product of its surface brightness $I_{j}$ and its apparent angular
size $d\Omega_{j}$.  But, for the face-on observer, each copy of the
flash ($j=0,1,2,\ldots$) has the same surface brightness
$I=\int_{0}^{\infty}I_{\nu}d\nu$, where $I_{\nu}$ is the specific
intensity.  [To see this, first note that $I_{\nu}/\nu^{3}$ is the
same in all Lorentz frames and conserved along a photon geodesic
\cite{MTW}.  Next note that the ratio $\nu_{0}/\nu_{e}$ between the
observed frequency of a photon ($\nu_{o}$) and the frequency it had in
the rest frame of the equatorial circularly-orbiting emitter
($\nu_{e}$) depends on $L_{z}$, but not on $Q$
\cite{CunninghamBardeen, Chandrasekhar}; so, for our face-on observer,
who only receives photons with $L_{z}=0$, the ratio is
$j$-independent.  In other words, there is no {\it relative} redshift
between the various copies of the flash received by the face-on
observer.  Since $I_{\nu}$ was isotropic in the emitter's rest frame,
$I_{\nu}$ and $I$ are also $j$-independent.]  Thus, $\mu_{j}$ is just
the ratio $d\Omega_{j}/d\Omega_{0}$ between the apparent size of the
$j$th flash ($d\Omega_{j}$) and the $0$th flash ($d\Omega_{0}$), which
may be calculated, given $Q_{j}(r_{\ast},a)$, as explained in
\cite{CunninghamBardeen, Chandrasekhar}.  See Figures 2 and 3.
\begin{figure}
  \begin{center}
    \includegraphics[width=3.1in]{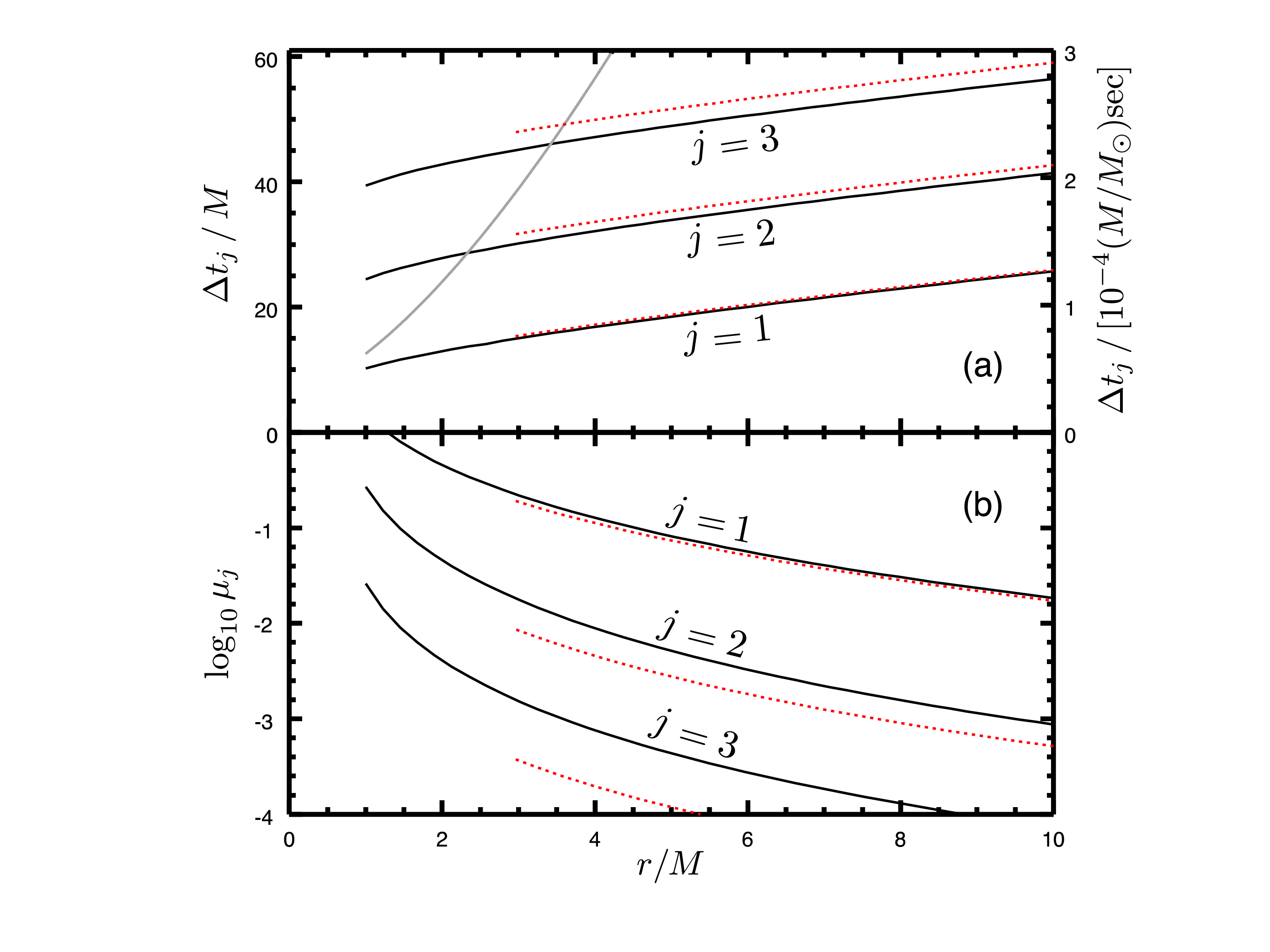}
  \end{center}
  \caption{Time delay $\Delta t_{j}(r_{\ast},a)$ and magnification
    $\mu_{j}(r_{\ast},a)$ of the $j=1,2,3$ paths relative to the $j=0$
    path, as a function of emission radius $r_{\ast}>r_{{\rm ph,eq}}$,
    for holes that are non-spinning ($a/M=0$, red dotted curves) and
    maximally spinning ($a/M=1$, black solid curves).  The solid grey
    curve in Panel a shows the orbital period (for $a/M=1$); again
    note that it is considerably longer than the blink separation.}
\label{afixedFig}
\end{figure}

Next, instead of a flash, let the emission have arbitrary (perhaps
unknown or stochastic) time variation.  If the $j=0$ photons reach the
face-on observer with light curve $L_{0}(t)$, then the full light
curve, including light loops, is $L(t)\!=\!\sum_{j=0}^{\infty}
\mu_{j}L_{0}(t\!-\!\Delta t_{j})$, where $\mu_{0}=1$ and $\Delta
t_{0}=0$; and if the emission is characterized by autocorrelation
function $\xi_{0}(T)\!\equiv\langle L_{0}(t)L_{0}(t-T)\rangle$, then
the observed autocorrelation function, including light loops, is
$\xi(T)\!\equiv\!\langle L(t)L(t-T)\rangle =\sum_{j,k=0}^{\infty}
\mu_{j}\mu_{k}\xi_{0}[T+(\Delta t_{k}-\Delta t_{j})]$.  Given a
promising astronomical source, these formulae for $L(t)$ and $\xi(T)$
correspond to two strategies to search for blinking (see Fig.~4).  (i)
Given (theoretical or empirical) information about the emitted light
curve ($\propto L_{0}(t)$), one can construct a family of blinking
light curves $L(t,r_{\ast},a,M)$ that may be correlated/fit to the
data, much as the LIGO experiment uses ``matched filtering'' to search
its noisy data for predicted gravitational waveforms.  (ii)
Alternatively, we can search for blinking in the auto-correlation
function $\xi(T)$.  This is better for sources that exhibit continuous
and random variability, rather than short well-separated bursts; and
it has the advantage that $\xi(T)$ needn't be measured on a
flare-by-flare basis -- rather, one can accumulate better statistics
over time ({\it e.g.}\ over many flares, or many observations).

The symmetry of the face-on configuration makes the blinking more
robust and independent of certain astrophysical and observational
details, in two ways.  First, since $\Delta t_{j}$ and $\mu_{j}$
depend on the radius $r_{\ast}$, but {\it not} on the azimuthal angle
$\varphi_{\ast}$ of the emission, the face-on light curve is sensitive
to the {\it total} emission from the equatorial ring of radius
$r_{\ast}$, not its $\varphi_{\ast}$ profile.  Second, there is no
\emph{relative} redshift between the various paths ($j=0,1,2,\ldots$),
so each blink $L_{j}(t)$ is a shifted copy of the primary $L_{0}(t)$;
to compute the effect, we don't need to know the frequency spectrum of
the source, or the frequency band of the detector.  Farther from the
face-on view, the time delays, magnifications and redshifts of the
various paths are increasingly $\varphi_{\ast}$ dependent, and the
blinking features in $\xi(T)$ are increasingly smeared out.

Refs.~\cite{CunninghamBardeen, BroderickLoeb} consider a source that
circularly orbits in the equatorial plane near a black hole and emits
with luminosity that is constant (or long-lived relative to the
orbital period), and calculate how the observed light curve $L(t)$
oscillates with the orbital period.  (In \cite{CunninghamBardeen} the
source is a star; in \cite{BroderickLoeb} it is a hotspot orbiting in
the accretion disk.)  This oscillation (which we call ``time-dependent
lensing'' or ``TDL'') is complementary to our blinking signal in
several respects.  TDL is due to the source's $\varphi_{\ast}$-motion
and $\varphi_{\ast}$-localization, not its intrinsic variability; by
constrast, blinking is due to the source's intrinsic variability, not
its $\varphi_{\ast}$-motion or $\varphi_{\ast}$-localization.  In the
face-on configuration, where we have argued that blinking is optimal,
TDL vanishes; and in the edge-on configuration, where TDL is
strongest, blinking is smeared out [in $\xi(T)$, not $L(t)$].  If the
observer is sufficiently face-on, blinking dominates over TDL; if the
observer is sufficiently non-face-on (and the source's emission is
sufficiently constant and $\varphi_{\ast}$-localized), TDL dominates
over blinking.  Observers who are more ``edge-on'' may (in some cases,
and at some frequencies) find the TDL signal obscured by dust;
observers who are more ``face-on'' may (in some cases and at some
frequencies) find the blinking signal obscured by a jet.

\begin{figure}
  \begin{center}
    \includegraphics[width=3.1in]{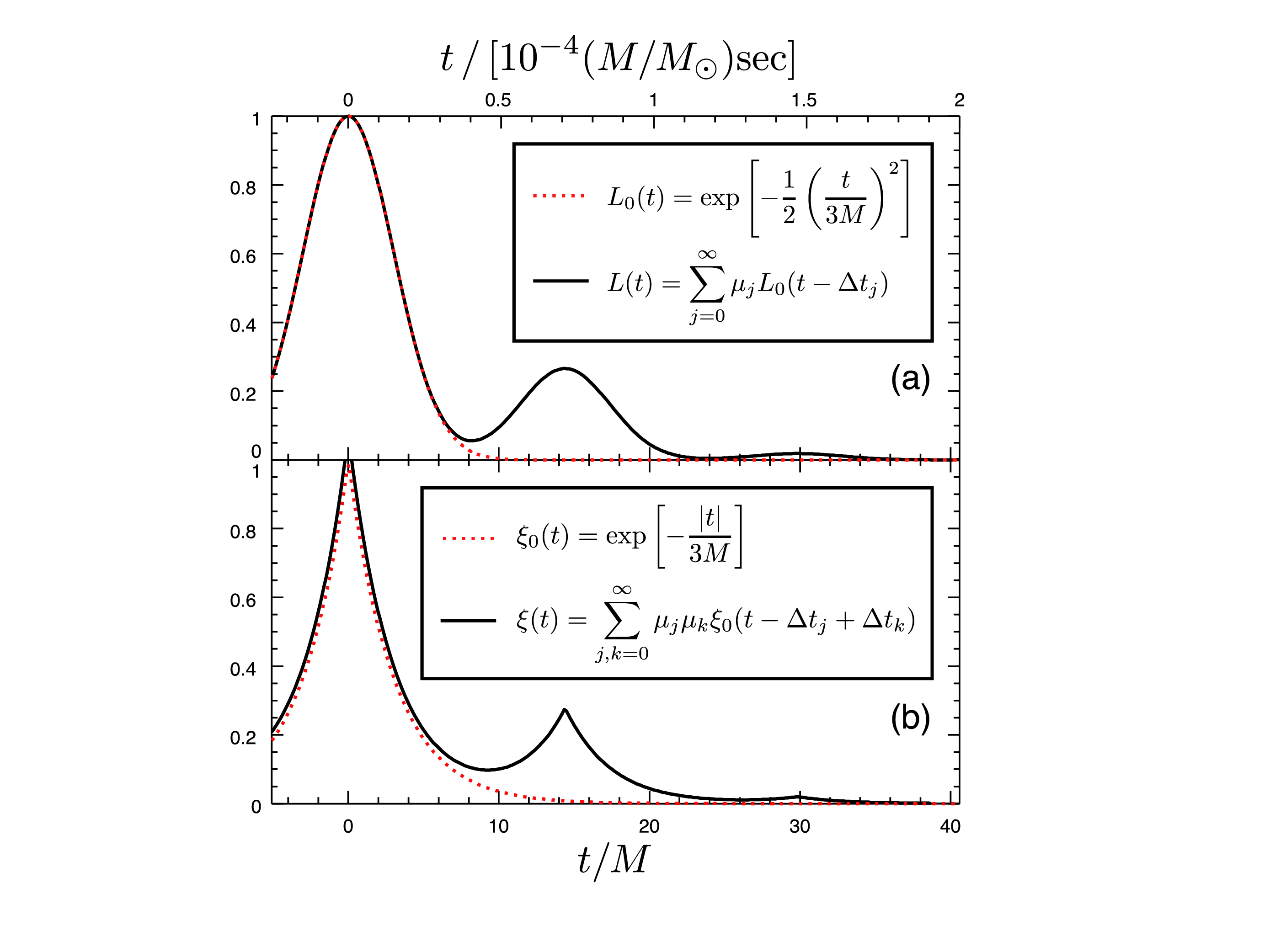}
  \end{center}
  \caption{For $a=a_{{\rm crit}}\approx0.853$ and $r_{\ast}=r_{{\rm
        isco}}$: (a) example light curve $L(t)$ and (b) example
    autocorrelation function $\xi(T)$.}
  \label{exampleFig}
\end{figure}
In calculating the blinking signal, we have assumed that the region
near the hole, where the rays propagate, is optically thin.  For many
systems, this assumption should be valid, at least in certain
wavelength ranges.  One also expects the optical depth to drop sharply
inside the ISCO \cite{PageThorne}; and since the highly bent rays that
ultimately reach the face-on observer do most of their bending very
near the polar ($L_{z}=0$) photon ring, whose radius is
\begin{equation}
  \label{r_cr}
  r_{{\rm ph,pol}}=1\!+\!2(1\!-\!\frac{1}{3}a^{2})^{1/2}
  {\rm cos}\left[\frac{1}{3}{\rm arccos}\frac{1\!-\!a^{2}}
    {(1\!-\!\frac{1}{3}a^{2})^{3/2}}\right]
\end{equation}
there is a critical value of the spin, $a_{{\rm crit}}\approx0.853$,
at which $r_{{\rm isco}}=r_{{\rm ph,pol}}$ (Fig.~2).  For $a<a_{{\rm
    crit}}$ ($a>a_{{\rm crit}}$), the $j\geq1$ light loops intersect
the equatorial plane inside (outside) the ISCO, where one can be more
(less) confident that the accretion flow is optically thin.  If the
optical depth along a given path, $\tau_{j}(r_{\ast},a)$, is not
negligible, we should make the replacement $\mu_{j}(r_{\ast},a)\to{\rm
  exp}[\tau_{0}(r_{\ast},a)-\tau_{j}(r_{\ast},a)]\mu_{j}(r_{\ast},a)$.

Many issues deserve further consideration.  Our analysis should be
extended to non-face-on configurations, and more realistic models of
the emission and optical depth.  Early on, we evoked a picture in
which the emission comes from the inner accretion disk; but our
analysis applies to more general systems, and it is worth thinking
broadly about the possibilities.  Which astronomical systems might
display blinking?  Certain stellar mass, intermediate mass, or
supermassive black hole systems?  Which observational frequency bands
and techniques are most promising?  Could a signal already be lurking
in any existing ({\it e.g.}\ radio, optical, x-ray, or gamma-ray) data
sets?  Might gravitational wave detectors help us to locate suitable
black hole systems?  It may even be worth mentioning that the blinking
effect is not restricted to electromagnetic emission: if a supernova
explodes near a black hole, we might see blinking in its neutrino
signal; or if two (stellar-mass) compact objects merge near a
(supermassive or intermediate-mass) black hole, we might see blinking
in their gravitational wave signal.  Is black hole blinking detectable?
We hope this paper will encourage further consideration of this
important question.

\acknowledgments We are grateful to Avery Broderick, Chris Hirata,
Mike Kesden, Norm Murray and Chris Thompson for valuable feedback.  LB
thanks the CIFAR JFA for support.


\begin{thebibliography}{99}
  
\bibitem{Darwin} C.~Darwin, Proc.\ Roy.\ Soc.\ London {\bf A249}, 180
  (1959).
  
\bibitem{CunninghamBardeen} C.T.~Cunningham and J.M.~Bardeen,
  Ap.\ J.\ {\bf 173}, 137 (1972); Ap.\ J.\ {\bf 183},
  237 (1973).

\bibitem{HolzWheeler} D.~Holz and J.~Wheeler, Astrophys.\ J.\ {\bf
    578}, 330 (2002).
  
\bibitem{BozzaEtAl} V.~Bozza and L.~Mancini, Ap.\ J.\ {\bf 627},
  790 (2005); V.~Bozza, S.C.~Novati and L.~Mancini, arXiv:0711.0750.
  
\bibitem{BroderickLoeb} A.~Broderick and A.~Loeb, MNRAS {\bf 363}, 353
  (2005); Ap.\ J.\ {\bf 636}, L109 (2006); MNRAS {\bf 367}, 905
  (2006); S.S.~Doeleman, V.L.~Fish, A.E.~Broderick, A.~Loeb and
  A.E.E.~Rogers, Ap.\ J.\ {\bf 695}, 59 (2009).

\bibitem{GravitationalLensing} P.~Schneider, J.~Ehlers and E.E.~Falco,
  {\it Gravitational Lenses}, Springer (1999).

\bibitem{PageThorne} D.~Page and K.~Thorne, Astrophys.\ J.\
  {\bf 191}, 499 (1974).

\bibitem{Bardeen:1970} J.M.~Bardeen, Nature {\bf 226}, 64 (1970).
  
\bibitem{BardeenPetterson} J.M.~Bardeen and J.A.~Petterson,
  Ap.\ J.\ {\bf 195}, L65 (1975).

\bibitem{Kerr} R.P.~Kerr, Phys.\ Rev.\ Lett.\ 1963.

\bibitem{Carter} B.~Carter, Phys.\ Rev.\ D {\bf 174}, 1559 (1968).

\bibitem{BardeenPressTeukolsky} J.M.~Bardeen, W.H.~Press and
  S.A.~Teukolsky, Ap.\ J.\ {\bf 178}, 347 (1972).

\bibitem{Chandrasekhar} S.~Chandrasekhar, {\it The Mathematical Theory
    of Black Holes}, Oxford University Press (1983).

\bibitem{MTW} C.W.~Misner, K.S.~Thorne and J.A.~Wheeler, {\it
    Gravitation}, W.H.~Freeman and Company (1973).

\end{thebibliography}
\end{document}